\documentclass[reqno]{amsart}
\usepackage{mathrsfs,amssymb,amsmath,amsfonts,amsxtra}
\usepackage{graphicx}
\usepackage{verbatim}
\newcommand{\href}[2]{#2}  

\begin{document}
\sloppypar \sloppy
\title[Comments on Schulz]{Comments on a paper by B.Schulz about Bell's inequalities}

\author{I. Schmelzer}
\thanks{Akeleiweg 7, 12487 Berlin, Germany, ilja.schmelzer@gmail.com}



\begin{abstract}
Schulz claims to have constructed an actively local stochastic theory which violates Bell's inequality. This is false.
\end{abstract}
\maketitle

\newcommand{\pd}{\partial} 
\newcommand{\ud}{\mathrm{d}} 

\newcommand{\B}{\mbox{$\mathbb{Z}_2$}} 
\newcommand{\Z}{\mbox{$\mathbb{Z}$}}
\newcommand{\R}{\mbox{$\mathbb{R}$}}
\newcommand{\C}{\mbox{$\mathbb{C}$}}
\renewcommand{\i}{{p}}
\renewcommand{\l}{{\lambda}}
\renewcommand{\L}{{\Lambda}}
\renewcommand{\o}{{\omega}}
\renewcommand{\O}{{\Omega}}

\newtheorem{theorem}{Theorem}
\newtheorem{axiom}{Axiom}
\newtheorem{definition}[axiom]{Definition}

\vspace{1cm}

In the conclusions of his paper \cite{Schulz}, Schulz claims:
\begin{quote}
Specially designed stochastic processes can violate Bell's inequality at two separated locations without any contact at all. We have mathematically analysed, what such a behaviour really does imply for a stochastic theory, and we have constructed a corresponding one.
\end{quote}

This is false. Schulz constructs a variant of Nelson's stochastic interpretation proposed by Fritsche and Haugk \cite{Fritsche}, which he introduces with the words
\begin{quote}
In contrast to Nelson's original contribution \ldots, the model of Fritsche and Haugk is an actively local
theory with a non-markovian stochastic process.
\end{quote}
The violation of Bell's inequality would be a consequence of the recovery of full quantum theory (as in original stochastic mechanics) in his construction. But \cite{Schulz}, Eq. (88) gives the remarkable restriction 
\begin{equation}
\begin{split}
 \vec{v}^{NA(B)} &\equiv \left(\vec{v}_1^{A(B)}(\vec{r}_1,t),\ldots,\vec{v}_N^{A(B)}(\vec{r}_N,t)\right),\\
 \vec{u}^{NA(B)} &\equiv \left(\vec{u}_1^{A(B)}(\vec{r}_1,t),\ldots,\vec{u}_N^{A(B)}(\vec{r}_N,t)\right)
\end{split} 
\end{equation} 
that each $\vec{v}_i$, $\vec{u}_i$ depends only on $\vec{r}_i$. Using the natural many-particle generalization of Ref. \cite{Schulz}, Eq. (66) and following
\begin{equation}\label{v}
\psi(\vec{r},t) \equiv \pm \sqrt{\rho(\vec{r},t)}e^{i\phi(\vec{r},t)}, \qquad \vec{v} \equiv \frac{\hbar}{m_0}\vec{\nabla}\phi(\vec{r},t),
\end{equation}
which requires only to interpret $\vec{r}$ as an element of $\R^{3N}$, one obtains
\begin{equation}
\phi(\vec{r}) = \sum_k \phi_k(\vec{r}_k).
\end{equation}
Combining it with the natural many-particle generalization of Ref. \cite{Schulz}, Eq. (59)
\begin{equation}\label{u}
 \vec{u}^{A(B)} \equiv \pm \nu \vec{\nabla}\ln \left(\rho^{A(B)}(\vec{r},t)/\rho_0\right),
\end{equation}
one gets
\begin{equation}
 \ln \rho(\vec{r}) = \sum_k f_k(\vec{r}_k).
\end{equation}
As a consequence the wave function splits into a simple product state
\begin{equation}
\psi(\vec{r}) = \prod_k e^{f_k(\vec{r}_k)+i\phi_k(\vec{r}_k)} = \prod_k \psi_k(\vec{r}_k).
\end{equation} 

Given the otherwise quite unmotivated section ``4.2 Quantum states in superpositions'' of \cite{Schulz} (in pilot wave theory or Nelson's stochastics there is no need to consider superpositions separately), one may guess that Schulz tries to overcome this restriction there. Unfortunately the whole construction, especially the considerations from (70) to (75) of \cite{Schulz}, are nonsensical. And even if one would accept all this, one can use his equations
\begin{equation}
 \vec{v}_\uparrow=\vec{v}_\downarrow\equiv\vec{v}; \quad \vec{u}_\uparrow=\vec{u}_\downarrow\equiv\vec{u}
\end{equation}
(in the text after Ref. \cite{Schulz}, Eq.  (83)) to obtain (modulo the integration constants in our \eqref{u}, \eqref{v} which become irrelevant constant factors) that
\begin{equation}
 \psi_\uparrow=\psi_\downarrow=\psi,
\end{equation}
so that no nontrivial superposition has been obtained.\footnote{Another way to see that this consideration fails (which remains valid also for the ``corrected version'' in \cite{SchulzReply}) is to consider the coefficients $a$, $b$ in the superposition $a\psi_\uparrow+b\psi_\downarrow$: They appear there as arbitrary coefficients, without any relation to the construction itself. But nontrivial superpositions with different $a$, $b$ are in general physically different, thus, cannot be obtained from a construction which does not even depend on $a$, $b$. Corrections proposed in \cite{SchulzReply2} don't solve this problem.} Thus, full quantum theory is not recovered. No direct proof that the construction nonetheless allows to violate Bell's inequality is given.\footnote{In \cite{SchulzReply} Schulz draws a surprising conclusion: ``Thereby the theory given in my article contradicts the assumption of passive locality. Hence it violates Bell’s theorem.''. So it seems that Schulz thinks that he has given such a proof. But does he really think that to prove a violation of Bell's inequality it is sufficient to prove that one assumption of one version of a proof does not hold? In \cite{SchulzReply2}, the same error is repeated: ``Deterministic passive locality is required to hold in order to derive Bell’s inequalities \ldots. As a result, it is obvious, that Bell’s inequalities can be violated by the modified Fritsche-Haugk theory.''}

Even more, one can even directly prove that the construction cannot violate Bell's inequality. Schulz himself describes the way how to do this at Ref. \cite{Schulz}, p. 31:
\begin{quote}
The only constraint on the random forces was that they should have a Gaussian distribution. Now consider the following mechanism: A real random number generator sets a value $\lambda_j\in\R$ for each $j$-th particle pair at preparation stage. This value $\lambda_j$ is similar to the hidden parameter in Bell's original work [\footnote{Here Ref. \cite{Bell}.}], but we define it to be independent for each entangled particle pair generated by the source.
\end{quote}
Now, Bell's $\lambda$ will be in general also different for each entangled pair, thus, using a different $\lambda_j$ for each pair does not pose a problem if we want to fit into Bell's framework. Schulz 
continues:
\begin{quote}
In our model, $\lambda_j$ serves as the starting value for two pseudo random number generators of the same type.
\end{quote}
Pseudo-random number generators nicely fit into Bell's original scheme, where all one needs are well-defined deterministic functions $A(a,\lambda)$, $B(b,\lambda)$. So what could prevent us from applying Bell's original proof? The following (see Ref. \cite{Schulz}, p. 31) sounds gramatically like a counterargument:
\begin{quote}
At first sight, one may think that our procedure would set up a deterministic passively local theory. Indeed, for $\vec{F}^{Brown}_{j1}(t,\lambda_j)$ and $\vec{F}^{Brown}_{j2}(t,\lambda_j)$  there is a parameter $\lambda_j$ at preparation stage which determines all later outcomes of these forces. However, one does not have access to the outcomes of the Gaussian distributed random forces in the probability space, but only to the two spin values \ldots
\end{quote}
But whatever the aim of this remark, it does not prevent us from applying Bell's theorem, which does not require that the $\l\in\L$ (usually named ``hidden variables'') should be accessible in any way to anybody. What is important is that the observable results $A, B$ can be obtained as deterministic functions $A(a,\l)$, $B(b,\l)$ for some set of ``beables'' $\L$. This seems unquestioned.

So is there any argument why one cannot apply Bell's theorem? Schulz, implicitly suggesting some (non-existing) lack of mathematical rigour in Bell's proof, writes (see Ref. \cite{Schulz}, p. 2)
\begin{quote}
In 1985, Nelson tried to analyse Bell's theorem with full mathematical rigour. \ldots Bell's definition of ``locally causal'' \ldots could be divided into two separate conditions \ldots
\end{quote}
One of them is named ``passive locality'' (also named ``output independence'', but it is essentially simply completeness in the EPR sense). It requires the independence of the (stochastic) functions $A$, $B$,
\begin{equation}
P(A,B|a,b,\l)=P(A|a,b,\l)P(B|a,b,\l),
\end{equation}
given the parameters $a,b$ controlled by the experimenters and some additional parameters (hidden variables, beables or common causes) $\l\in\L$. Then he claims (Ref. \cite{Schulz}, p. 32):
\begin{quote}
We will illustrate below that deterministic passive locality can be easily violated, even though the outcomes of the time dependent random force may be predetermined by a hidden parameter for each entangled particle pair.
\end{quote}
But if ``passive locality'' holds depends on which conditional probabilities are considered: $P(A,B|a,b,\l)$ may not factorize into $P(A|a,b,\l)P(B|a,b,\l)$ for one space $\L\ni\l$ but factorize for another (greater) space $\L$. If, in particular, $\L$ contains all hidden parameters, so that $A(a,b,\l)$, $B(a,b,\l)$ become deterministic functions, the corresponding probability distribution factorizes trivially:
\begin{equation}\label{factor}
P(A,B|a,b,\l)=\delta(A-A(a,b,\l))\delta(B-B(a,b,\l)).
\end{equation}
Schulz uses an incomplete $\L$ (his $\mathcal{F}_{jS}$) which does not contain the $\l_j$ of the pseudo-random number generators. Therefore it would be nonsensical to expect that the probability distribution factorizes, and, indeed, it does not. But in a proof of Bell's inequality for his construction one would use a larger, complete $\L$ containing also all the $\l_j$ of the random number generators, so that everything is deterministic and ``passive locality'' holds trivially because of Eq. \eqref{factor}. Of course, Schulz is free to consider whatever conditional probabilities he likes. But other people are also free to consider other probabilities in their proof that for his construction Bell's inequality holds.

Thus, Schulz has not presented a valid reason why Bell's theorem is not applicable to his construction. One can apply it and prove in this way that his construction is unable to violate Bell's inequality. Thus, the claimed result of the paper \cite{Schulz} is wrong.

But, maybe, the very idea to save relativity and realism with stochastic theories which don't violate ``active locality'' (parameter independence) but only ``passive causality'' (output independence) deserves future research? I don't think so. Any theory of this type faces the good old EPR argument that it is simply incomplete --- it doesn't explain the observed correlations, because an explanation is the presentation of a cause for the correlation (a sufficiently large $\L$) such that once the cause is taken into account, the correlation disappears.

In general, stochastic functions do not seem to me very promising for this purpose. If stochastic functions $A(\l,a)$, $B(\l,b)$ have a parameter-independent probability measure $d\nu(\o)$ on some sample space $\O$, a simple redefinition $\L\to\L'\cong \L\times\O$, $d\rho(l)\to d\rho'(\l')=d\rho(\l)d\nu(\o)$ makes Bell's original theorem for deterministic functions $A(\l,a)$, $B(\l,b)$ applicable. Allowing parameter dependence $d\nu(\o|a,b)$ leads effectively to solipsism --- the pure observable results $d\rho(A,B|a,b)$ themself would already define a complete ``realistic model''. Thus, Kolmogorovian stochastics does not help. And a non-Kolmogorovian generalization will automatically face the argument that it violates realism, that it does not give a realistic explanation.

On the other hand, there is no reason to give up realism. In the opinion of the author, the road to realism is to accept a hidden preferred frame, as in pilot wave theory and Nelsonian stochastics. This is not only compatible with all existing evidence, but also allows new possibilities like a condensed matter interpretation for fundamental fields \cite{clm}.

\end{document}